\input epsf
\newfam\scrfam
\batchmode\font\tenscr=rsfs10 \errorstopmode
\ifx\tenscr\nullfont
        \message{rsfs script font not available. Replacing with calligraphic.}
        \def\scr{\cal}
\else   
        \font\sevenscr=rsfs7
        \font\fivescr=rsfs5
        \skewchar\tenscr='177 \skewchar\sevenscr='177 \skewchar\fivescr='177
        \textfont\scrfam=\tenscr \scriptfont\scrfam=\sevenscr
        \scriptscriptfont\scrfam=\fivescr
        \def\scr{\fam\scrfam}
        \def\cal{\scr}
\fi
\catcode`\@=11
\newfam\frakfam
\batchmode\font\tenfrak=eufm10 \errorstopmode
\ifx\tenfrak\nullfont
        \message{eufm font not available. Replacing with italic.}
        
\else
	
	\font\sevenfrak=eufm7 \font\fivefrak=eufm5
        
	\textfont\frakfam=\tenfrak
	\scriptfont\frakfam=\sevenfrak \scriptscriptfont\frakfam=\fivefrak
	
\fi
\catcode`\@=\active
\newfam\msbfam
\batchmode\font\twelvemsb=msbm10 scaled\magstep1 \errorstopmode
\ifx\twelvemsb\nullfont\def\Bbb{\bf}
        
	\font\eightbbb=cmb10 at 8pt
	\message{Blackboard bold not available. Replacing with boldface.}
\else   \catcode`\@=11
        \font\tenmsb=msbm10 \font\sevenmsb=msbm7 \font\fivemsb=msbm5
        \textfont\msbfam=\tenmsb
        \scriptfont\msbfam=\sevenmsb \scriptscriptfont\msbfam=\fivemsb
        \def\Bbb{\relax\expandafter\Bbb@}
        \def\Bbb@#1{{\Bbb@@{#1}}}
        \def\Bbb@@#1{\fam\msbfam\relax#1}
        \catcode`\@=\active
	
	\font\eightbbb=msbm8
\fi
        \font\fivemi=cmmi5
        \font\sixmi=cmmi6
        \font\eightrm=cmr8              \def\xrm{\eightrm}
        \font\eightbf=cmbx8             \def\xbf{\eightbf}
        \font\eightit=cmti10 at 8pt     \def\xit{\eightit}
                
        \font\eighttt=cmtt8             
        \font\eightcp=cmcsc8
                      \def\xold{\eighti}
        \font\eightmi=cmmi8
                     \def\xbold{\eightib}
                       \def\old{\teni}
        \font\tencp=cmcsc10

        \font\twelvecp=cmcsc10 scaled\magstep1
        
        \font\sixrm=cmr6
        \font\fiverm=cmr5

        \font\eightsy=cmsy8
        \font\sixsy=cmsy6
        \font\eightsl=cmsl8
        \font\sixbf=cmbx6

	 at10pt	
	\font\twelvehelvbold=phvb at12pt
	 at14pt
	\font\sixteenhelvbold=phvb at16pt

\def\noblackbox{\overfullrule=0pt}
\noblackbox

\def\eightpoint{
\def\rm{\fam0\eightrm}
\textfont0=\eightrm \scriptfont0=\sixrm \scriptscriptfont0=\fiverm
\textfont1=\eightmi  \scriptfont1=\sixmi  \scriptscriptfont1=\fivemi
\textfont2=\eightsy \scriptfont2=\sixsy \scriptscriptfont2=\fivesy
\textfont3=\tenex   \scriptfont3=\tenex \scriptscriptfont3=\tenex
\textfont\itfam=\eightit \def\it{\fam\itfam\eightit}
\textfont\slfam=\eightsl \def\sl{\fam\slfam\eightsl}
\textfont\ttfam=\eighttt \def\tt{\fam\ttfam\eighttt}
\textfont\bffam=\eightbf \scriptfont\bffam=\sixbf 
                         \scriptscriptfont\bffam=\fivebf
                         \def\bf{\fam\bffam\eightbf}
\normalbaselineskip=10pt}

\newtoks\headtext
\headline={\ifnum\pageno=1\hfill\else
	\ifodd\pageno{\eightcp\the\headtext}{ }\dotfill{ }{\old\folio}
	\else{\old\folio}{ }\dotfill{ }{\eightcp\the\headtext}\fi
	\fi}
\def\makeheadline{\vbox to 0pt{\vss\noindent\the\headline\break
\hbox to\hsize{\hfill}}
        \vskip2\baselineskip}
\newcount\infootnote
\infootnote=0
\newcount\footnotecount
\footnotecount=1
\def\foot#1{\infootnote=1
\footnote{${}^{\the\footnotecount}$}{\vtop{\baselineskip=.75\baselineskip
\advance\hsize by
-\parindent{\eightpoint\rm\hskip-\parindent
#1}\hfill\vskip\parskip}}\infootnote=0\global\advance\footnotecount by
1}
\newcount\refcount
\refcount=1
\newwrite\refwrite
\def\oldsize{\ifnum\infootnote=1\xold\else\old\fi}
\def\ref#1#2{
	\def#1{{{\oldsize\the\refcount}}\ifnum\the\refcount=1\immediate\openout\refwrite=\jobname.refs\fi\immediate\write\refwrite{\item{[{\xold\the\refcount}]} 
	#2\hfill\par\vskip-2pt}\xdef#1{{\noexpand\oldsize\the\refcount}}\global\advance\refcount by 1}
	}
\def\refout{\eightpoint\catcode`\@=11
        \xrm\immediate\closeout\refwrite
        \vskip2\baselineskip
        {\noindent\twelvecp References}\hfill\vskip\baselineskip
        \baselineskip=.75\baselineskip
        \input\jobname.refs
        \baselineskip=4\baselineskip \divide\baselineskip by 3
        \catcode`\@=\active\rm}

\def\skipref#1{\hbox to15pt{\phantom{#1}\hfill}\hskip-15pt}

\def\hepth#1{\href{http://xxx.lanl.gov/abs/hep-th/#1}{arXiv:\allowbreak
hep-th\slash{\xold#1}}}

\def\matharx#1{\href{http://xxx.lanl.gov/abs/math/#1}{arXiv:math/{\xold#1}}}
\def\arxiv#1#2{\href{http://arxiv.org/abs/#1.#2}{arXiv:\allowbreak
{\xold#1}.{\xold#2}} [hep-th]} 
\def\arxivmdg#1#2{\href{http://arxiv.org/abs/#1.#2}{arXiv:\allowbreak
{\xold#1}.{\xold#2}} [math.DG]} 
\def\jhep#1#2#3#4{\href{http://jhep.sissa.it/stdsearch?paper=#2\%28#3\%29#4}{J. High Energy Phys. {\xbold #1#2} ({\xold#3}) {\xold#4}}}

\def\CQG#1#2#3{Class. Quantum Grav. {\xbold#1} ({\xold#2}) {\xold#3}}
\def\FP#1#2#3{Fortsch. Phys. {\xbold#1} ({\xold#2}) {\xold#3}}

\def\IJMPCS#1#2#3{Int. J. Mod. Phys. Conf. Ser. {\xbf A}{\xbold#1} ({\xold#2}) {\xold#3}}

\def\JMP#1#2#3{J. Math. Phys. {\xbold#1} ({\xold#2}) {\xold#3}}
\def\JPA#1#2#3{J. Phys. {\xbf A}{\xbold#1} ({\xold#2}) {\xold#3}}
\def\LMP#1#2#3{Lett. Math. Phys. {\xbold#1} ({\xold#2}) {\xold#3}}
\def\MPLA#1#2#3{Mod. Phys. Lett. {\xbf A}{\xbold#1} ({\xold#2}) {\xold#3}}

\def\NPB#1#2#3{Nucl. Phys. {\xbf B}{\xbold#1} ({\xold#2}) {\xold#3}}

\def\PLB#1#2#3{Phys. Lett. {\xbf B}{\xbold#1} ({\xold#2}) {\xold#3}}
\def\PR#1#2#3{Phys. Rept. {\xbold#1} ({\xold#2}) {\xold#3}}
\def\PRD#1#2#3{Phys. Rev. {\xbf D}{\xbold#1} ({\xold#2}) {\xold#3}}
\def\PRL#1#2#3{Phys. Rev. Lett. {\xbold#1} ({\xold#2}) {\xold#3}}

\def\TMP#1#2#3{Theor. Math. Phys. {\xbold#1} ({\xold#2}) {\xold#3}}
\def\TMF#1#2#3{Teor. Mat. Fiz. {\xbold#1} ({\xold#2}) {\xold#3}}
\newcount\sectioncount
\sectioncount=0
\def\section#1#2{\global\eqcount=0
	\global\subsectioncount=0
        \global\advance\sectioncount by 1
	\ifnum\sectioncount>1
	        \vskip2\baselineskip
	\fi
\line{\twelvecp\the\sectioncount. #2\hfill}
       \vskip.5\baselineskip\noindent
        \xdef#1{{\old\the\sectioncount}}}
\newcount\subsectioncount
\def\subsection#1#2{\global\advance\subsectioncount by 1
\vskip.75\baselineskip\noindent\line{\tencp\the\sectioncount.\the\subsectioncount. #2\hfill}\nobreak\vskip.4\baselineskip\nobreak\noindent\xdef#1{{\old\the\sectioncount}.{\old\the\subsectioncount}}}
\def\immediatesubsection#1#2{\global\advance\subsectioncount by 1
\vskip-\baselineskip\noindent
\line{\tencp\the\sectioncount.\the\subsectioncount. #2\hfill}
	\vskip.5\baselineskip\noindent
	\xdef#1{{\old\the\sectioncount}.{\old\the\subsectioncount}}}
\newcount\subsubsectioncount
\def\subsubsection#1#2{\global\advance\subsubsectioncount by 1
\vskip.75\baselineskip\noindent\line{\tencp\the\sectioncount.\the\subsectioncount.\the\subsubsectioncount. #2\hfill}\nobreak\vskip.4\baselineskip\nobreak\noindent\xdef#1{{\old\the\sectioncount}.{\old\the\subsectioncount}.{\old\the\subsubsectioncount}}}
\newcount\appendixcount
\appendixcount=0
\def\appendix#1{\global\eqcount=0
        \global\advance\appendixcount by 1
        \vskip2\baselineskip\noindent
        \ifnum\the\appendixcount=1
        \hbox{\twelvecp Appendix A: #1\hfill}\vskip\baselineskip\noindent\fi
    \ifnum\the\appendixcount=2
        \hbox{\twelvecp Appendix B: #1\hfill}\vskip\baselineskip\noindent\fi
    \ifnum\the\appendixcount=3
        \hbox{\twelvecp Appendix C: #1\hfill}\vskip\baselineskip\noindent\fi}
\def\acknowledgements{\vskip2\baselineskip\noindent
        \underbar{\it Acknowledgements:}\ }
\newcount\eqcount
\eqcount=0
\def\Eqn#1{\global\advance\eqcount by 1
\ifnum\the\sectioncount=0
	\xdef#1{{\noexpand\oldsize\the\eqcount}}
	\eqno({\oldstyle\the\eqcount})
\else
        \ifnum\the\appendixcount=0
\xdef#1{{\noexpand\oldsize\the\sectioncount}.{\noexpand\oldsize\the\eqcount}}
                \eqno({\oldstyle\the\sectioncount}.{\oldstyle\the\eqcount})\fi
        \ifnum\the\appendixcount=1
	        \xdef#1{{\noexpand\oldstyle A}.{\noexpand\oldstyle\the\eqcount}}
                \eqno({\oldstyle A}.{\oldstyle\the\eqcount})\fi
        \ifnum\the\appendixcount=2
	        \xdef#1{{\noexpand\oldstyle B}.{\noexpand\oldstyle\the\eqcount}}
                \eqno({\oldstyle B}.{\oldstyle\the\eqcount})\fi
        \ifnum\the\appendixcount=3
	        \xdef#1{{\noexpand\oldstyle C}.{\noexpand\oldstyle\the\eqcount}}
                \eqno({\oldstyle C}.{\oldstyle\the\eqcount})\fi
\fi}
\def\eqn{\global\advance\eqcount by 1
\ifnum\the\sectioncount=0
	\eqno({\oldstyle\the\eqcount})
\else
        \ifnum\the\appendixcount=0
                \eqno({\oldstyle\the\sectioncount}.{\oldstyle\the\eqcount})\fi
        \ifnum\the\appendixcount=1
                \eqno({\oldstyle A}.{\oldstyle\the\eqcount})\fi
        \ifnum\the\appendixcount=2
                \eqno({\oldstyle B}.{\oldstyle\the\eqcount})\fi
        \ifnum\the\appendixcount=3
                \eqno({\oldstyle C}.{\oldstyle\the\eqcount})\fi
\fi}
\def\multi{\global\advance\eqcount by 1}
\def\multieqn#1{({\oldstyle\the\sectioncount}.{\oldstyle\the\eqcount}\hbox{#1})}
\def\multiEqn#1#2{\xdef#1{{\oldstyle\the\sectioncount}.{\old\the\eqcount}#2}
        ({\oldstyle\the\sectioncount}.{\oldstyle\the\eqcount}\hbox{#2})}
\def\multiEqnAll#1{\xdef#1{{\oldstyle\the\sectioncount}.{\old\the\eqcount}}}
\newcount\tablecount
\tablecount=0
\def\Table#1#2{\global\advance\tablecount by 1
       \xdef#1{\the\tablecount}
       \vskip2\parskip
       \centerline{\it Table \the\tablecount: #2}
       \vskip2\parskip}
\newtoks\url
\def\Href#1#2{\catcode`\#=12\url={#1}\catcode`\#=\active#2}
\def\href#1#2{{#2}}

\parskip=3.5pt plus .3pt minus .3pt
\baselineskip=14pt plus .1pt minus .05pt
\lineskip=.5pt plus .05pt minus .05pt
\lineskiplimit=.5pt
\abovedisplayskip=18pt plus 4pt minus 2pt
\belowdisplayskip=\abovedisplayskip
\hsize=14cm
\vsize=19cm
\hoffset=1.5cm
\voffset=1.8cm
\frenchspacing
\footline={}
\raggedbottom

\newskip\origparindent
\origparindent=\parindent

\def\*{\partial}
\def\punkt{\,\,.}
\def\komma{\,\,,}

\def\={\!=\!}
\def\small#1{{\hbox{$#1$}}}

\def\fraction#1{\small{1\over#1}}
\def\fr{\fraction}
\def\Fraction#1#2{\small{#1\over#2}}
\def\Fr{\Fraction}

\def\eg{{\it e.g.}}

\def\ie{{\it i.e.}}

\def\nlni{\hfill\break}

\def\RR{{\Bbb R}}




\def\textfrac#1#2{\raise .45ex\hbox{\the\scriptfont0 #1}\nobreak\hskip-1pt/\hskip-1pt\hbox{\the\scriptfont0 #2}}

\def\LL{{\cal L}}
\def\leftbr{[\![}
\def\rightbr{]\!]}


\def\frac{\Fr}

\def\mathbb{\Bbb}



\def\LL{{\cal L}}
\def\leftbr{[\![}
\def\rightbr{]\!]}

\def\LL{{\cal L}}
\def\leftbr{[\![}
\def\rightbr{]\!]}


\def\old{\rm}
\def\xold{\xrm}
\def\xbold{\xbf}
\def\oldstyle{}


\ref\HohmZwiebachLarge{O. Hohm and B. Zwiebach, {\xit ``Large gauge
transformations in double field theory''}, \jhep{13}{02}{2013}{075}
[\arxiv{1207}{4198}].} 

\ref\AschieriEtAl{P. Aschieri, I. Bakovi\'c, B. Jur\v co and
P. Schupp, {\xit ``Noncommutative gerbes and deformation
quantization''}, \hepth{0206101}.}

\ref\BermanCederwallKleinschmidtThompson{D.S. Berman, M. Cederwall,
A. Kleinschmidt and D.C. Thompson, {\xit ``The gauge structure of
generalised diffeomorphisms''}, \jhep{13}{01}{2013}{64} [\arxiv{1208}{5884}].}

\ref\CederwallUfoldbranes{M. Cederwall, {\xit ``M-branes on U-folds''},
in proceedings of 7th International Workshop ``Supersymmetries and
Quantum Symmetries'' Dubna, 2007 [\arxiv{0712}{4287}].}

\ref\BermanPerryGen{D.S. Berman and M.J. Perry, {\xit ``Generalised
geometry and M-theory''}, \jhep{11}{06}{2011}{074} [\arxiv{1008}{1763}].}    

\ref\BermanMusaevThompson{D.S. Berman, E.T. Musaev and D.C. Thompson,
{\xit ``Duality invariant M-theory: gaugings via Scherk--Schwarz
reduction}, \jhep{12}{10}{2012}{174} [\arxiv{1208}{0020}].}

\ref\UdualityMembranes{V. Bengtsson, M. Cederwall, H. Larsson and
B.E.W. Nilsson, {\xit ``U-duality covariant
membranes''}, \jhep{05}{02}{2005}{020} [\hepth{0406223}].}

\ref\ObersPiolineU{N.A. Obers and B. Pioline, {\xit ``U-duality and M-theory''},
\PR{318}{1999}{113}, 
\nlni [\hepth{9809039}].}

\ref\BermanGodazgarPerry{D.S. Berman, H. Godazgar and M.J. Perry,
{\xit ``SO(5,5) duality in M-theory and generalized geometry''},
\PLB{700}{2011}{65} [\arxiv{1103}{5733}].} 

\ref\BermanMusaevPerry{D.S. Berman, E.T. Musaev and M.J. Perry,
{\xit ``Boundary terms in generalized geometry and doubled field theory''},
\PLB{706}{2011}{228} [\arxiv{1110}{97}].} 

\ref\BermanGodazgarGodazgarPerry{D.S. Berman, H. Godazgar, M. Godazgar  
and M.J. Perry,
{\xit ``The local symmetries of M-theory and their formulation in
generalised geometry''}, \jhep{12}{01}{2012}{012}
[\arxiv{1110}{3930}].} 

\ref\BermanGodazgarPerryWest{D.S. Berman, H. Godazgar, M.J. Perry and
P. West,
{\xit ``Duality invariant actions and generalised geometry''}, 
\jhep{12}{02}{2012}{108} [\arxiv{1111}{0459}].} 

\ref\CoimbraStricklandWaldram{A. Coimbra, C. Strickland-Constable and
D. Waldram, {\xit ``$E_{d(d)}\times\hbox{\eightbbb R}^+$ generalised geometry,
connections and M theory'' }, \arxiv{1112}{3989}.} 

\ref\CremmerPopeI{E. Cremmer, B. Julia, H. L\"u and C.N. Pope,
{\xit ``Dualisation of dualities. I.''}, \NPB{523}{1998}{73} [\hepth{9710119}].}

\ref\HullT{C.M. Hull, {\xit ``A geometry for non-geometric string
backgrounds''}, \jhep{05}{10}{2005}{065} [\hepth{0406102}].}

\ref\HullM{C.M. Hull, {\xit ``Generalised geometry for M-theory''},
\jhep{07}{07}{2007}{079} [\hepth{0701203}].}

\ref\HullZwiebachII{C. Hull and B. Zwiebach, {\xit ``Double field
theory''}, \jhep{09}{09}{2009}{99} [\arxiv{0904}{4664}].}

\ref\HullZwiebach{C. Hull and B. Zwiebach, {\xit ``The gauge algebra
of double field theory and Courant brackets''},
\jhep{09}{09}{2009}{90} [\arxiv{0908}{1792}].}

\ref\HullDoubled{C.M. Hull, {\xit ``Doubled geometry and
T-folds''}, \jhep{07}{07}{2007}{080}
[\hepth{0605149}].}

\ref\HullTownsend{C.M. Hull and P.K. Townsend, {\xit ``Unity of
superstring dualities''}, \NPB{438}{1995}{109} [\hepth{9410167}].}

\ref\PalmkvistHierarchy{J. Palmkvist, {\xit ``Tensor hierarchies,
Borcherds algebras and $E_{11}$''}, \jhep{12}{02}{2012}{066}
[\arxiv{1110}{4892}].} 

\ref\deWitNicolaiSamtleben{B. de Wit, H. Nicolai and H. Samtleben,
{\xit ``Gauged supergravities, tensor hierarchies, and M-theory''},
\jhep{02}{08}{2008}{044} [\arxiv{0801}{1294}].}

\ref\deWitSamtleben{B. de Wit and H. Samtleben,
{\xit ``The end of the $p$-form hierarchy''},
\jhep{08}{08}{2008}{015} [\arxiv{0805}{4767}].}

\ref\CederwallJordanMech{M.~Cederwall, {\xit ``Jordan algebra
dynamics''}, \PLB{210}{1988}{169}.} 

\ref\BerkovitsNekrasovCharacter{N. Berkovits and N. Nekrasov, {\xit
    ``The character of pure spinors''}, \LMP{74}{2005}{75}
  [\hepth{0503075}].}

\ref\HitchinLectures{N. Hitchin, {``\xit Lectures on generalized
geometry''}, \arxiv{1010}{2526}.}

\ref\KoepsellNicolaiSamtleben{K. Koepsell, H. Nicolai and
H. Samtleben, {\xit ``On the Yangian $[Y(e_8)]$ quantum symmetry of
maximal supergravity in two dimensions''}, \jhep{99}{04}{1999}{023}
[\hepth{9903111}].}

\ref\HohmHullZwiebachI{O. Hohm, C.M. Hull and B. Zwiebach, {\xit ``Background
independent action for double field
theory''}, \jhep{10}{07}{2010}{016} [\arxiv{1003}{5027}].}

\ref\HohmHullZwiebachII{O. Hohm, C.M. Hull and B. Zwiebach, {\xit
``Generalized metric formulation of double field theory''},
\jhep{10}{08}{2010}{008} [\arxiv{1006}{4823}].} 

\ref\HohmZwiebach{O. Hohm and B. Zwiebach, {\xit ``On the Riemann
tensor in double field theory''}, \jhep{12}{05}{2012}{126}
[\arxiv{1112}{5296}].} 

\ref\WestEEleven{P. West, {\xit ``$E_{11}$ and M theory''},
\CQG{18}{2001}{4443} [\hepth{0104081}].}

\ref\AndriotLarforsLustPatalong{D. Andriot, M. Larfors, D. L\"ust and
P. Patalong, {\xit ``A ten-dimensional action for non-geometric
fluxes''}, \jhep{11}{09}{2011}{134} [\arxiv{1106}{4015}].}

\ref\AndriotHohmLarforsLustPatalongI{D. Andriot, O. Hohm, M. Larfors,
D. L\"ust and 
P. Patalong, {\xit ``A geometric action for non-geometric
fluxes''}, \PRL{108}{2012}{261602} [\arxiv{1202}{3060}].}

\ref\AndriotHohmLarforsLustPatalongII{D. Andriot, O. Hohm, M. Larfors,
D. L\"ust and 
P. Patalong, {\xit ``Non-geometric fluxes in supergravity and double
field theory''}, \FP{60}{2012}{1150} [\arxiv{1204}{1979}].}

\ref\DamourHenneauxNicolai{T. Damour, M. Henneaux and H. Nicolai,
{\xit ``Cosmological billiards''}, \CQG{20}{2003}{R145} [\hepth{0212256}].}

\ref\DamourNicolai{T. Damour and H. Nicolai, 
{\xit ``Symmetries, singularities and the de-emergence of space''},
\arxiv{0705}{2643}.}

\ref\EHTP{F. Englert, L. Houart, A. Taormina and P. West,
{\xit ``The symmetry of M theories''},
\jhep{03}{09}{2003}{020}2003 [\hepth{0304206}].}

\ref\PachecoWaldram{P.P. Pacheco and D. Waldram, {\xit ``M-theory,
exceptional generalised geometry and superpotentials''},
\jhep{08}{09}{2008}{123} [\arxiv{0804}{1362}].}

\ref\DamourHenneauxNicolaiII{T. Damour, M. Henneaux and H. Nicolai,
{\xit ``$E_{10}$ and a 'small tension expansion' of M theory''},
\PRL{89}{2002}{221601} [\hepth{0207267}].}

\ref\KleinschmidtNicolai{A. Kleinschmidt and H. Nicolai, {\xit
``$E_{10}$ and $SO(9,9)$ invariant supergravity''},
\jhep{04}{07}{2004}{041} [\hepth{0407101}].}

\ref\WestII{P.C. West, {\xit ``$E_{11}$, $SL(32)$ and central charges''},
\PLB{575}{2003}{333} [\hepth{0307098}].}

\ref\KleinschmidtWest{A. Kleinschmidt and P.C. West, {\xit
``Representations of $G^{+++}$ and the r\^ole of space-time''},
\jhep{04}{02}{2004}{033} [\hepth{0312247}].}

\ref\WestIII{P.C. West, {\xit ``$E_{11}$ origin of brane charges and
U-duality multiplets''}, \jhep{04}{08}{2004}{052} [\hepth{0406150}].}

\ref\PiolineWaldron{B. Pioline and A. Waldron, {\xit ``The automorphic
membrane''}, \jhep{04}{06}{2004}{009} [\hepth{0404018}].}

\ref\WestBPS{P.C. West, {\xit ``Generalised BPS conditions''},
\arxiv{1208}{3397}.}

\ref\PalmkvistBorcherds{J. Palmkvist, {\xit ``Borcherds and Kac--Moody
  extensions of simple finite-dimensional Lie algebras''}, \arxiv{1203}{5107}.}

\ref\ParkSuh{J.-H. Park and Y. Suh, {\xit ``U-geometry: SL(5)''},
\arxiv{1302}{1652}.} 

\ref\CederwallMinimalExcMult{M. Cederwall, {\xit ``Non-gravitational 
exceptional supermultiplets''}, 
\jhep{13}{07}{2013}{025}
[\arxiv{1302}{6737}].} 

\ref\PalmkvistDual{J. Palmkvist, {\xit ``The tensor hierarchy
algebra''}, \arxiv{1305}{0018}.}

\ref\CederwallPalmkvistSerre{M. Cederwall and J. Palmkvist, {\xit
    ``Serre relations, constraints and partition functions''}, to appear.}

\ref\CoimbraStricklandWaldramII{A. Coimbra, C. Strickland-Constable and
D. Waldram, {\xit ``Supergravity as generalised geometry II:
$E_{d(d)}\times\hbox{\eightbbb R}^+$ and M theory''}, \arxiv{1212}{1586}.}  

\ref\HohmZwiebachGeometry{O. Hohm and B. Zwiebach, {\xit ``Towards an
invariant geometry of double field theory''}, \arxiv{1212}{1736}.} 

\ref\JeonLeeParkRR{I. Jeon, K. Lee and J.-H. Park, {\xit
``Ramond--Ramond cohomology and O(D,D) T-duality''},
\jhep{12}{09}{2012}{079} [\arxiv{1206}{3478}].} 

\ref\PureSGI{M. Cederwall, {\xit ``Towards a manifestly supersymmetric
    action for D=11 supergravity''}, \jhep{10}{01}{2010}{117}
    [\arxiv{0912}{1814}].}  

\ref\PureSGII{M. Cederwall, 
{\xit ``D=11 supergravity with manifest supersymmetry''},
    \MPLA{25}{2010}{3201} [\arxiv{1001}{0112}].}

\ref\CremmerLuPopeStelle{E. Cremmer, H. L\"u, C.N. Pope and
K.S. Stelle, {\xit ``Spectrum-generating symmetries for BPS solitons''},
\NPB{520}{1998}{132} [\hepth{9707207}].}

\ref\JeonLeeParkI{I. Jeon, K. Lee and J.-H. Park, {\xit ``Differential
geometry with a projection: Application to double field theory''},
\jhep{11}{04}{2011}{014} [\arxiv{1011}{1324}].}

\ref\JeonLeeParkII{I. Jeon, K. Lee and J.-H. Park, {\xit ``Stringy
differential geometry, beyond Riemann''}, 
\PRD{84}{2011}{044022} [\arxiv{1105}{6294}].}

\ref\JeonLeeParkIII{I. Jeon, K. Lee and J.-H. Park, {\xit
``Supersymmetric double field theory: stringy reformulation of supergravity''},
\PRD{85}{2012}{081501} [\arxiv{1112}{0069}].}

\ref\HohmKwak{O. Hohm and S.K. Kwak, {\xit ``$N=1$ supersymmetric
double field theory''}, \jhep{12}{03}{2012}{080} [\arxiv{1111}{7293}].}

\ref\HohmKwakFrame{O. Hohm and S.K. Kwak, {\xit ``Frame-like geometry
of double field theory''}, \JPA{44}{2011}{085404} [\arxiv{1011}{4101}].}

\ref\HohmKwakZwiebachI{O. Hohm, S.K. Kwak and B. Zwiebach, {\xit
``Unification of type II strings and T-duality''},
\PRL{107}{2011}{171603} [\arxiv{1106}{5452}].}

\ref\HohmKwakZwiebachII{O. Hohm, S.K. Kwak and B. Zwiebach, {\xit
``Double field theory of type II strings''}, \jhep{11}{09}{2011}{013}
[\arxiv{1107}{0008}].} 

\ref\Hillmann{C. Hillmann, {\xit ``Generalized $E_{7(7)}$ coset
dynamics and $D=11$ supergravity''}, \jhep{09}{03}{2009}{135}
[\arxiv{0901}{1581}].}

\ref\AldazabalGranaMarquesRosabal{G. Aldazabal, M. Gra\~na,
D. Marqu\'es and J.A. Rosabal, {\xit ``Extended geometry and gauged
maximal supergravity''}, \arxiv{1302}{5419}.}

\ref\DuffDualityString{M.J. Duff, {\xit ``Duality rotations in string
theory''}, \NPB{335}{1990}{610}.}

\ref\DuffDualityMembrane{M.J. Duff and J.X. Lu, {\xit ``Duality
rotations in membrane theory''}, \NPB{347}{1990}{394}.}

\ref\HohmLustZwiebach{O. Hohm, D. L\"ust and B. Zwiebach, {\xit ``The
spacetime of double field theory: Review, remarks and outlook''},
\arxiv{1309}{2977}.} 

\ref\HullEtAlGerbes{C.M. Hull, U. Lindstr\"om, M. Ro\v cek, R. von
Unge and M. Zabzine, {\xit ``Generalized K\"ahler geometry and
gerbes''}, \jhep{09}{10}{2009}{062} [\arxiv{0811}{3615}].}

\ref\Kikuchi{T.~Kikuchi, T.~Okada and Y.~Sakatani,
  {\xit ``Rotating string in doubled geometry with generalized isometries''},
  \PRD{86}{2012}{046001}
  [\arxiv{1205}{5549}].}

\ref\KikuchiGL{  T.~Kikuchi, T.~Okada and Y.~Sakatani,
  {\xit ``Generalized Lie transported string in T-fold with
  generalized Killing vector''}, 
  \IJMPCS{21}{2013}{169}.}

\ref\Banks{T.~Banks and L.J.~Dixon,
  {\xit ``Constraints on string vacua with space-time supersymmetry''},
  \NPB{307}{1988}{93}.}
 
\ref\Park{J.-H.~Park,
  {\xit ``Comments on double field theory and diffeomorphisms''},
  \jhep{13}{06}{2013}{098}
  [\arxiv{1304}{5946}].}

\ref\ParkLeeGauge{K.~Lee and J.-H.~Park,
  {\xit ``Covariant action for a string in doubled yet gauged spacetime''}
  \arxiv{1307}{8377}.}

\ref\Borisov{A.B.~Borisov and V.I.~Ogievetsky,
  {\xit ``Theory of dynamical affine and conformal symmetries as
  gravity theory''}, 
  \TMP{21}{1975}{1179}
   [\TMF{21}{1974}{329}].}

\ref\Tseytlin{A.A.~Tseytlin,
  {\xit ``Duality symmetric closed string theory and interacting
  chiral scalars''}, 
  \NPB{350}{1991}{395}.}


\ref\Shigemori{J.~de Boer and M.~Shigemori,
  {\xit ``Exotic branes and non-geometric backgrounds''},
  \PRL{104}{2010}{251603}
  [\arxiv{1004}{2521}].}

\ref\ShigemoriEB{J.~de Boer and M.~Shigemori,
  {\xit ``Exotic branes in string theory''},
  \PR{532}{2013}{65}
  [\arxiv{1209}{6056}].}

\ref\Hitchin{N.J.~Hitchin,
 {\xit ``Lectures on special Lagrangian submanifolds''},
  \matharx{9907034}.}

\ref\LustNA{R.~Blumenhagen, M.~Fuchs, D.~L\"ust and R.~Sun,
 {\xit ``Non-associative deformations of geometry in double field theory''},
  \arxiv{1312}{0719}.}

\ref\Blumenhagen{R.~Blumenhagen,
  {\xit ``Nonassociativity in string theory''},
  \arxiv{1112}{4611}.}

\ref\BlumenhagenAlgebroid{R. Blumenhagen, A. Deser, E. Plauschinn and
F. Rennecke, {\xit ``Non-geometric strings, symplectic gravity and
differential geometry of Lie algebroids ''}, 
\jhep{13}{02}{2013}{122} [\arxiv{1304}{2784}].}

\ref\BlumenhagenAlgebroidII{R. Blumenhagen, A. Deser, E. Plauschinn,
F. Rennecke and C. Schmid, {\xit ``The intriguing structure of
non-geometric frames in string theory''}, \arxiv{1304}{2784}.}

\ref\Bakas{I.~Bakas and D.~L\"ust,
  {\xit ``3-cocycles, non-associative star-products and the magnetic
  paradigm of R-flux string vacua''}, 
  \arxiv{1309}{3172}.}

\ref\Marques{G.~Aldazabal, D.~Marqu\'es and C.~N\'u\~nez,
  {\xit ``Double field theory: A pedagogical review''},
  \CQG{30}{2013}{163001}
  [\arxiv{1305}{1907}].}

\ref\BermanThompson{D.S.~Berman and D.C.~Thompson,
  {\xit``Duality symmetric string and M-theory''},
  \arxiv{1306}{2643}.}

\ref\SiegelI{W.~Siegel,
  {\xit ``Two vierbein formalism for string inspired axionic gravity''},
  \PRD{47}{1993}{5453}
  [\hepth{9302036}].}

\ref\SiegelII{ W.~Siegel,
  {\xit ``Superspace duality in low-energy superstrings''},
  \PRD{48}{1993}{2826}
  [\hepth{9305073}].}

\ref\SiegelIII{W.~Siegel,
  {\xit ``Manifest duality in low-energy superstrings''},
  in Berkeley 1993, Proceedings, Strings '93 353
  [\hepth{9308133}].}

\ref\Belov{D.M.~Belov, C.M.~Hull and R.~Minasian,
  {\xit ``T-duality, gerbes and loop spaces''},
  \arxiv{0710}{5151}.}

\ref\HullZwiebachCourant{C.~Hull and B.~Zwiebach,
  {\xit ``The gauge algebra of double field theory and Courant brackets''},
  \jhep{09}{09}{2009}{090}
  [\arxiv{0908}{1792}].}

\ref\Sugawara{S.~Kawai and Y.~Sugawara,
  {\xit ``Mirrorfolds with K3 fibrations''},
  \jhep{08}{02}{2008}{065}
  [\arxiv{0711}{1045}].}



\ref\Szabo{D. Mylonas, P. Schupp and R.J. Szabo,
 {\xit ``Membrane sigma-models and quantization of non-geometric flux
backgrounds"}
  \jhep{12}{09}{2012}{012}
  [\arxiv{1207}{0926}].}

\ref\PachecoWaldram{P.~P.~Pacheco and D.~Waldram,
  {\xit ``M-theory, exceptional generalised geometry and superpotentials''},
  \jhep{08}{09}{2008}{123}
  [\arxiv{0804}{1362}].}

\ref\CederwallI{M.~Cederwall, J.~Edlund and A.~Karlsson,
  {\xit ``Exceptional geometry and tensor fields''},
  \jhep{13}{07}{2013}{028}
  [\arxiv{1302}{6736}].}

\ref\CederwallII{ M.~Cederwall,
  {\xit ``Non-gravitational exceptional supermultiplets''},
  \jhep{13}{07}{2013}{025}
  [\arxiv{1302}{6737}].}

\ref\SambtlebenHohmI{O.~Hohm and H.~Samtleben,
  {\xit ``Exceptional field theory I: $E_{6(6)}$ covariant form of
  M-theory and type IIB''}, 
  \arxiv{1312}{0614}.}

\ref\SambtlebenHohmII{O.~Hohm and H.~Samtleben,
  {\xit ``Exceptional field theory II: $E_{7(7)}$''},
  \arxiv{1312}{4542}.}

\ref\Diego{G.~Aldazabal, M.~Gra\~na, D.~Marqu\'es and J.A.~Rosabal,
  {\xit ``The gauge structure of exceptional field theories and the
  tensor hierarchy''}, 
  \arxiv{1312}{4549}.}

\ref\Riccioni{F.~Riccioni and P.~C.~West,
  {\xit``E(11)-extended spacetime and gauged supergravities''},
  \jhep{08}{02}{2008}{039}
  [\arxiv{0712}{1795}].}

\ref\Rocen{A.~Roc\'en and P.~West,
  {\xit``E11, generalised space-time and IIA string theory: the R-R sector''},
  \arxiv{1012}{2744}.}

\ref\Duff{M.~J.~Duff,
  {\xit ``Duality rotations in string theory''},
  \NPB{335}{1990}{610}.}

\ref\VaismanI{I.~Vaisman,
  {\xit ``On the geometry of double field theory''},
  \JMP{53}{2012}{033509}
  [\arxivmdg{1203}{0836}].}

\ref\VaismanII{I.~Vaisman,
  {\xit ``Towards a double field theory on para-Hermitian manifolds''},
  \arxivmdg{1209}{0152}.}

\ref\VaismanIII{I.~Vaisman,
  {\xit ``Geometry on big-tangent manifolds''},
  \arxivmdg{1303}{0658}.}

\ref\BlairMalekPark{C.D.A. Blair, E. Malek and J.-H. Park,
{\xit ``M-theory and F-theory from a duality manifest action''},
\arxiv{1311}{5109}.}

\ref\StricklandConstableSubsectors{C. Strickland-Constable, 
{\xit ``Subsectors, Dynkin diagrams and new generalised geometries''},
\arxiv{1310}{4196}.}

\ref\BlairMalekRouth{C.D.A. Blair, E. Malek and A.J. Routh,
{\xit ``An O(D,D) invariant Hamiltonian action for the superstring''},
\arxiv{1308}{4829}.}

\ref\FreidelLeighMinic{L. Freidel, R.G. Leigh and Dj. Mini\'c,
{\xit ``Born reciprocity in string theory and the nature of spacetime''},
\arxiv{1307}{7080}.}

\ref\BermanBlairMalekPerryODD{D.S. Berman, C.D.A. Blair, E. Malek and
M.J. Perry, 
{\xit ``The $O_{D,D}$ geometry of string theory''},
\arxiv{1303}{6727}.} 

\ref\BermanCederwallPerry{D.S. Berman, M. Cederwall and M.J. Perry,
{\xit ``Global aspects of double geometry''}, \arxiv{1401}{1311}.}


\ref\Papadopoulos{G. Papadopoulos, {\xit ``Seeking the balance:
Patching double and exceptional field theories''}, \arxiv{1402}{2586}.}

\ref\BetzBlumenhagenLustRennecke{A. Betz, R. Blumenhagen, D. L\"ust
and F. Rennecke, {\xit ``A note on the CFT origin of the strong
constraint of DFT''}, \arxiv{1402}{1686}.}

\ref\Geissbuhler{D. Geissb\"uhler, {\xit ``Double field theory and N=4
gauged supergravity''}, \jhep{11}{11}{2011}{116} [\arxiv{1109}{4280}].}

\ref\AldazabalBaronMarquesNunez{G. Aldazabal, W. Baron, D. Marqu\'es
ans C. N\'u\~nez, {\xit ``The effective action of double field
theory''}, \jhep{11}{11}{2011}{052} [arxiv{1109}{0290}].}


\line{
\epsfysize=18mm
\epsffile{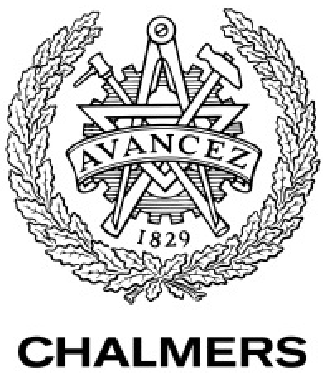}
\hfill}
\vskip-11mm

\line{\hfill Gothenburg preprint}
\line{\hfill February, {\old2014}}

\line{\hrulefill}

\headtext={Cederwall: 
``The geometry behind double geometry''}

\vfill
\vskip.5cm

\centerline{\sixteenhelvbold
The geometry behind double geometry}

\vfill

\vfill

\centerline{\twelvehelvbold Martin Cederwall}

\vfill
\vskip-1cm

\centerline{\it Dept. of Fundamental Physics}
\centerline{\it Chalmers University of Technology}
\centerline{\it SE 412 96 Gothenburg, Sweden}

\vfill

{\narrower\noindent \underbar{Abstract:}
Generalised diffeomorphisms in double field theory rely on an $O(d,d)$
structure defined on tangent space. 
We show that any (pseudo-)Riemannian metric
on the doubled space defines such a structure, in the sense that the
generalised diffeomorphisms defined using such a metric form an
algebra, provided a covariant section condition is
fulfilled. Consistent solutions of the section condition gives further
restrictions. The case
previously considered corresponds to a flat metric. The
construction makes it possible to apply double geometry to a larger
class of manifolds. Examples of curved defining metrics are given.
We also comment on the r\^ole of the defining
geometry for the symmetries of double field theory, and on the
continuation of the present construction to the U-duality setting. 
\smallskip}
\vfill

\font\xxtt=cmtt6

\vtop{\baselineskip=.6\baselineskip\xxtt
\line{\hrulefill}
\catcode`\@=11
\line{email: martin.cederwall@chalmers.se\hfill}
\catcode`\@=\active
}

\eject

\def\textfrac#1#2{\raise .45ex\hbox{\the\scriptfont0 #1}\nobreak\hskip-1pt/\hskip-1pt\hbox{\the\scriptfont0 #2}}

\noindent Generalised geometry 
(see refs. 
[\HullT\skipref\HullDoubled\skipref\Hitchin\skipref\HitchinLectures\skipref\SiegelI\skipref\HullM\skipref\HullZwiebach\skipref\HullZwiebachII\skipref\HohmHullZwiebachI\skipref\HohmHullZwiebachII\skipref\HohmZwiebach\skipref\HohmZwiebachGeometry\skipref\JeonLeeParkRR\skipref\JeonLeeParkI\skipref\JeonLeeParkII\skipref\JeonLeeParkIII\skipref\HohmKwak\skipref\HohmKwakFrame\skipref\HohmKwakZwiebachI\skipref\HohmKwakZwiebachII\skipref\ParkLeeGauge\skipref\BermanBlairMalekPerryODD\skipref\BermanPerryGen\skipref\BermanGodazgarPerry\skipref\BermanGodazgarGodazgarPerry\skipref\BermanGodazgarPerryWest\skipref\CoimbraStricklandWaldram\skipref\CoimbraStricklandWaldramII\skipref\BermanCederwallKleinschmidtThompson\skipref\ParkSuh\skipref\CederwallI\skipref\CederwallMinimalExcMult\skipref\AldazabalGranaMarquesRosabal\skipref\SambtlebenHohmI\skipref\SambtlebenHohmII\skipref\Diego\skipref\BlairMalekPark-\StricklandConstableSubsectors])
has arisen as a means to geometrise
duality symmetries. By using extended space-times, reduced to the
physical ones by a section condition, the local symmetries of gravity
and tensor fields unite in the generalised diffeomorphisms. This
permits not only a more symmetric formulation of the massless degrees
of string theory or M-theory, but also a situation where, for certain
backgrounds, duality symmetries are understood from a manifestly
geometric perspective, and where it is possible to move beyond
strictly geometric backgrounds.

Recent work 
[\HohmZwiebachLarge\skipref\Park\skipref\BermanCederwallPerry\skipref\HohmLustZwiebach-\Papadopoulos] 
has shed more light on global questions of generalised
geometry. 
In particular, the global questions about the structure of extended
manifolds have been asked and partially answered, although there are
remaining issues. This is closely related to the original main purpose of
the programme, namely to make the appearance of duality symmetries as
manifest as possible. It is important to note that these should not be
built into the formalism as global symmetries, but rather arise as
``generalised isometries'', special symmetries arising in special
(\eg\ toroidal) backgrounds, just as isometries arise in in ordinary
geometry.
It is by now known
[\HullZwiebach,\HullZwiebachII,\BermanCederwallPerry,\Papadopoulos]
that the full duality group can not be obtained this way, at least not
with the present formalism and level of understanding. As we will see,
one way forward
may be to include ordinary diffeomorphisms.

The extended manifolds have however so far been restricted to manifolds
equipped with a globally defined flat metric. This is because
the definition of the generalised diffeomorphism transformations
involves such a flat metric $\eta_{MN}$. This metric defines an
$O(d,d)$ structure on the extended space. In the following, we will
examine to what extent the defining metric can be chosen
differently. If one wants to consider not only generalised
diffeomorphisms (under which the defining metric is invariant), but
also ordinary diffeomorphisms, this becomes 
necessary, already for the case of the flat metric.

The usual definition of the double diffeomorphisms contains a
parameter $\xi^M$ on the doubled space. Fields transform under double
diffeomorphisms so that, in addition to the translation generated by
$\xi=\xi^M\*_M$, they are rotated by an $so(d,d)$ transformation
generated by $a-a^t$, where $a_M{}^N=\*_M\xi^N$. Here, the transpose
is defined using a constant metric $\eta_{MN}$, invariant under
$O(d,d)\subset GL(2d)$, such that
$(a^t)_M{}^N=\eta_{MP}\eta^{NQ}\*_Q\xi^P$.
Acting on a covector (which is equivalent to a vector, using $\eta$),
the standard form of a double diffeomorphism thus becomes
$$
\LL_\xi V=(\xi+a-a^t)V=(L_\xi-a^t)V\punkt\Eqn\FlatDoubleDiff
$$
It is then straightforward to verify that, when all fields, including
the transformation parameters, obey the section condition 
$$
\eta^{MN}\*_M\otimes\*_N=0\Eqn\FlatSectCond
$$ 
(the ``$\otimes$'' notation meaning that
the two derivative may act on the same field or any pair), the commutator
of two double diffeomorphisms is again a double diffeomorphism:
$$
\eqalign{
&[\LL_\xi,\LL_\eta]=\LL_{\leftbr\xi,\eta\rightbr}\komma\cr
&\hbox{where}\quad
\leftbr\xi,\eta\rightbr=\fr2(\LL_\xi\eta-\LL_\eta\xi)
\punkt\cr
}\Eqn\DoubleDiffAlgebra
$$
In ref. [\BermanCederwallPerry], this was elucidated by the
observation that 
$$
\leftbr\xi,\eta\rightbr=[\xi,\eta]+\chi_{\xi,\eta}\komma\eqn
$$
where 
$\chi^M_{\xi,\eta}=\fr2(-\xi^N\*^M\eta_N+\eta^N\*^M\xi_N)$ 
is a non-translating parameter, \ie, one for
which $\chi^M\*_M=0$ using the section condition. Therefore, 
$\LL_{\chi_{\xi,\eta}}=\Delta_{\xi,\eta}$ is a specific local
$so(d,d)$ transformation, which also turns out to be nilpotent. 
Using also $b_M{}^N=\*_M\eta^N$, the explicit form of $\Delta$ is
$$
\Delta_{\xi,\eta}=-ab^t+ba^t\punkt\eqn
$$
The
commutator can thus also be written
$$
[\LL_\xi,\LL_\eta]=\LL_{[\xi,\eta]}+\Delta_{\xi,\eta}\punkt\Eqn\LCommWithDelta
$$
This observation was used in ref. [\BermanCederwallPerry] to explain
the abelian gerbe structure encoded in double diffeomorphisms.

We thus observe that the double diffeomorphisms rely on the existence
of a flat metric $\eta_{MN}$. 
This implies no restriction locally, but limits the choice
of double manifolds to those globally allowing such a metric
structure\foot{This of course only implies if one insists on the
possibility of applying ordinary diffeomorphisms.}. These of course include tori, which are of special interest
since they lead to the ordinary (discrete) T-duality. 
It seems to be of no immediate interest to introduce curvature
locally, since the defining metric is non-dynamical, but it may be
important to be able to include topologies that demand a non-flat
metric, or indeed also in the flat situation. 
This is the main subject of this paper.

A local $O(d,d)$ structure is induced by any metric $H_{MN}$.
We therefore ask the questions:
{\it To what extent is it possible to use a (non-flat) metric $H$ on
the double space? What are the restrictions on such a metric implied
by the existence of an algebra of double diffeomorphisms?}
As we will see, any (pseudo-)Riemannian metric (of signature $(d,d)$) is
algebraically allowed, and further restriction follow only from the
existence of solutions to the section condition.

We would like to stress that the introduction of the metric $H$ has
nothing to do with equipping the double manifold with a generalised
metric, containing the metric and $B$-field on a subspace obtained by
solving the section condition. The metric $H$ is thought of as an
ordinary metric on the double space, whose purpose is to define a
local $O(d,d)$ structure. It will define an ordinary torsion-free
affine connection $\Gamma$ and a Riemann tensor $R$. 

The Ansatz we will use for the double diffeomorphisms is the natural
one that reduces to eq. (\FlatDoubleDiff) when $H=\eta$. In order to
obtain covariance, we should then use covariant derivatives $D$,
containing $\Gamma$, throughout, and use the covariantly constant
metric $H$ to raise and lower indices (\ie, to define the transpose of
a matrix). 
Consider therefore a transformation defined as
$$
\LL_\xi V_M=\xi^ND_NV_M+(a-a^t)_M{}^NV_N
=(L_\xi-a^t)_M{}^NV_N\komma\Eqn\HDoubleDiff
$$
where $a_M{}^N=D_M\xi^N$ and $(a^t)_M{}^N=H_{MP}H^{NQ}a_Q{}^P$. As
usual,
the connection terms in the Lie derivative cancel, and the choice of
defining metric is only reflected in the last term.
It is clear that $H$ itself is conserved by such a transformation
(extended to tensors), since it is covariantly constant and
$$
\LL_\xi H_{MN}=2(a-a^t)_{(M}{}^PH_{N)P}=0\punkt\eqn
$$
In order to check the algebra of these generalised diffeomorphisms, we
also need a section condition, which will be the natural
generalisation of eq. (\FlatSectCond), namely
$$
H^{MN}D_M\otimes D_N=0\punkt\eqn
$$
Note that there is no need of a section condition involving the
metric, since it is covariantly constant.

Let us now commute two transformations of the type
(\HDoubleDiff). Noting that 
$$
[\LL_\xi,\LL_\eta]=[L_\xi-a^t,L_\eta-b^t]\eqn
$$
(where of course also $b$ is defined with the covariant derivative, 
$b_M{}^N=D_M\eta^N$), it becomes clear that any obstruction containing
curvature will reside in terms containing $a^t$ or $b^t$.
A careful calculation, now keeping track of the order of covariant
derivatives, yields
$$
\eqalign{
&([\LL_\xi,\LL_\eta]-\LL_{[\xi,\eta]})_M{}^N\cr
&\qquad=-[a,b^t]_M{}^N+[b,a^t]_M{}^N+\xi^P[D^N,D_P]\eta_M-\eta^P[D^N,D_P]\xi_M\cr
&\qquad=(-ab^t+ba^t)_M{}^N+2R^N{}_{PMQ}\xi^{[P}\eta^{Q]}\punkt\cr
}\Eqn\CommutingLs
$$
In the second step, the section condition has been used in the form
$a^tb=0$ etc. The question now is whether this remainder can be written
as a non-translating transformation $\LL_\chi$ as in the case of flat
metric.
Consider a parameter 
$\chi^M_{\xi,\eta}=\fr2(-\xi^ND^M\eta_N+\eta^ND^M\xi_N)$. Then
$$
\eqalign{
(\Delta_{\xi,\eta})_M{}^N=(\LL_\chi)_M{}^N&=(-ab^t+ba^t)_M{}^N
-\fr2\xi^P[D_M,D^N]\eta_P+\fr2\eta^P[D_M,D^N]\xi_P\cr
&=(-ab^t+ba^t)_M{}^N-R_M{}^N{}_{PQ}\xi^P\eta^Q\punkt\cr
}\Eqn\ChiTransf
$$ 
Comparing eqs. (\CommutingLs) and (\ChiTransf), we see that they are
equal modulo the (vanishing) torsion Bianchi identity
$R_{[PQM]}{}^N=0$.

This shows the somewhat surprising result that there is no curvature
obstruction to the existence of an algebra of 
double diffeomorphisms. The metric $H$ defining the local $O(d,d)$
structure can be taken as any (pseudo-)Riemannian metric. The commutator is
still formally given by eq. (\DoubleDiffAlgebra) 
or eq. (\LCommWithDelta), although the
definition of the bracket $\leftbr\cdot,\cdot\rightbr$ is
metric-dependent, so our new algebras are most likely non-isomorphic
to the flat one.

Once we have established the formal closure of the algebra (strictly
speaking, algebroid) of double diffeomorphisms with the defining
metric $H$, it is important also to investigate possible solutions of
the section condition. Solving the section condition amounts to
finding a $d$-dimensional isotropic subspace of tangent space,
spanned by the $\tilde m$ directions in a split $X^M=(x^m,y^{\tilde m})$, such
that $D_{\tilde m}=0$ on all fields. Acting with further covariant
derivatives gives the integrability condition $R_{\tilde m NP}{}^Q=0$.
If this condition is fulfilled there will be a class of choices of
coordinates where
$D_{\tilde m}=\*_{\tilde m}$. 
Examples of such metrics are given by the pp-wave-like space-times
$$
ds^2=H_{MN}dX^MdX^N=h_{mn}(x)dx^mdx^n
+2\delta_{m\tilde m}dx^mdy^{\tilde m}
\komma\eqn
$$
with a set of $d$ light-like Killing vectors $\*_{\tilde m}$. 
It is not obvious to us whether this provides an exhaustive list of
allowed defining metrics. In any case, this class is general enough to
give room for any topology of a physical compactification space
(with coordinates $x^m$).

There should be analogous structures in exceptional extended geometry.
Unlike the case of doubled space, both the generalised diffeomorphisms
and the section condition involve a structure which is not a metric,
but a tensor defining an $E_{n(n)}\times\RR^+$ structure, the so called $Y$
tensor [\BermanCederwallKleinschmidtThompson]. 
The transformations look formally the same as in
eq. (\FlatDoubleDiff), but with $a^t$ replaced by 
$a^Y$, with components $(a^Y)_M{}^N=Y_{MP}{}^{QN}a_Q{}^P$, and the
``flat'' section condition reads $Y_{MN}{}^{PQ}\*_P\otimes\*_Q=0$. The
$Y$ tensor does not factorise into a product of a metric and its
inverse, so the structure sought for is not a metric
structure. Nevertheless, it should be possible to pursue a similar
investigation in these cases.

We would finally like to comment on the symmetries of double geometry
(the remarks apply to extended geometry in general).
It is known that the generalised diffeomorphisms are not general
enough to accommodate overlaps that would give truly non-geometric
solutions [\HullZwiebach,\HullZwiebachII,\BermanCederwallPerry,\Papadopoulos]. 
This is because of the section condition. Once a solution
to the section condition is chosen, it is preserved by the generalised
diffeomorphisms, which effectively prevents some duality
transformations, namely those that would act on the extended space in
way that changes the solution to the section condition (the section
condition itself is of course preserved). 
Therefore, the whole (discrete) T-duality group of some
compactification can not, with the present understanding, be
constructed as generalised isometries.
(This situation is by no means improved by the construction of the
present paper, rather the opposite, since the possibilities of
changing the section condition tend to be fewer.)

A proper understanding of the section condition, and of possible ways
to relax it (see \eg\ refs. 
[\Geissbuhler,\AldazabalBaronMarquesNunez,\BermanMusaevThompson,\BetzBlumenhagenLustRennecke]) 
is thus one of the key problems in extended geometry. 
A final solution to this problem will probably have to await a
formulation where the section condition is not applied ``by hand'',
but arises dynamically, as does the string theory level matching
condition. In the meantime, it is reasonable to expect
that in such a formulation, although the defining metric (or
exceptional structure) is not dynamical, its symmetries, which in the
case of double field theory are the isometries of $H$, can be included
as gauge symmetries. If this is the case, this will suffice to fill
out the T-duality group. 
We note that, while ordinary diffeomorphisms in general do not have
good commutators with the generalised diffeomorphisms $\LL_\xi$, since
they change the defining metric, isometries do. If $u^M$ is a Killing
vector of $H$ it is straightforward to check that
$[L_u,\LL_\xi]=\LL_{[u,\xi]}$ (the analogous statement for a finite
isometry is obviously true as well). Here, $u^M$ does not need to obey the
solution to the section condition, and it is only Killing vectors not
obeying it that generate transformations not contained in the $\LL$'s.
Hopefully, this way of constructing the full duality group can be a
first step in resolving the dilemma of obtaining duality symmetries
from extended field theories.

\acknowledgements The author would like to thank David Berman, Anna
Karlsson, George Papadopoulos and Malcolm Perry for discussions.

\vfill\eject

\refout

\end